\documentclass[jphysg,12pt]{article}
\usepackage{graphicx}
\usepackage{epsfig}
\begin{document}
\begin{center}
\Large {'Sling effect' in the development of atmospheric cascades
induced by primary cosmic ray nuclei}\\
\end{center}
\vspace {0.5cm}
\begin{center}
\footnote{E-mail address: a.d.erlykin@durham.ac.uk}
A.D.Erlykin~$^{(1,2)}$, A.W.Wolfendale$^{(2)}$
\end{center}
\begin{flushleft}
(1) {\em P. N. Lebedev Physical Institute, Moscow, Russia}\\
(2) {\em Department of Physics,University of Durham, Durham, UK}\\
\end{flushleft}

\begin{abstract}
The 'Sling effect' appears when a fragment of a projectile nucleus
emitted after its peripheral
 collision with a target nucleus is caused to rotate with high spin.
The spinning fragment has a deformed shape and looks like an
oblate ellipsoid. Due to the virtual non-compressibility of
nuclear matter, and the polarization of the spin in the plane
transverse to the input momentum of the projectile nucleus, such
an ellipsoid has a reduced mean interaction cross-section compared
with a non-spinning fragment which has a spherical shape. Purely
geometrical arguments dictate that such an ellipsoidal nucleus
should have additional fluctuations of cross-section even at a
fixed impact parameter dependent on the orientation angle between
the axis of the ellipsoid and the vector connecting the centers of
the projectile and target nucleus. The number of 'wounded
nucleons' in the projectile nucleus participating in the
interaction correlates strongly with the interaction
cross-section. All these effects lead to a non-exponential
attenuation of fragments and an increased probability for a
fragment to penetrate down to a larger depth in the absorber, than normal.

If the sling effect appears in the interaction of a primary cosmic
ray nucleus with nuclei in the atmosphere the induced atmospheric
cascade will have a slower attenuation, and thereby can help to
reduce some important inconsistencies in the interpretation of the existing
experimental data on extensive air showers observed in the lower
half of the atmosphere. The paper gives numerical estimates of the
sling effect.
\end{abstract}

\section{Introduction}

Information about nuclear interactions at energies higher than 2
PeV comes only from cosmic rays, viz. from the study of
atmospheric cascades initiated by primary cosmic ray protons and
heavier nuclei. Due to the largely unknown mass composition of the primary
cosmic rays the only criterium of the correctness of the chosen
interaction model, the mass composition and the adopted method of
analysis, is the consistency of the results on the primary mass
composition derived from different observables. A 'big leap'
forward in this direction has been made with the development of
the CORSIKA code for the simulation of extensive air showers
(~EAS~) which can incorporate and test different interaction
models \cite{Heck}. Using the results obtained with CORSIKA it was
shown that the QGSJET model gives in general the most consistent
description of the observed EAS characteristics and the mass
composition \cite{EW1,Ant1}.

However, deeper analysis of the EAS data collected by the KASCADE
experiment reveals that some serious inconsistencies still remain
\cite{Ulr1,Hau1}. There is a difference in the mass composition
derived from observables which either include or ignore $N_e$ -
the electron size of the shower, as well as between results
obtained from the ground-based measurements and from the
distribution of $X_{max}$ - the depth of the shower maximum in the
atmosphere measured by means of Cherenkov light, etc. There have
been different attempts to overcome these inconsistencies
\cite{Hoe1,Yako}. All of them agree that the atmospheric cascade
should be made more penetrative and propose different mechanisms
for that. We have also examined this problem and come to the
conclusion that the best way is to invoke a minor increase of the
energy fraction transferred to the electromagnetic component
(~K$_\gamma$ = 0.26~) with {\it an increase of the 'elongation
rate'} (the change in depth of maximum per decade of energy); the
value needed is about 71 gcm$^{-2}$. Both modifications are easier
to associate with nucleus-nucleus (~AA~) interactions than with
(~PA~) collisions \cite{EW2}. In the present paper we proceed
further in this direction.

\section{Sling effect}

\subsection{Fragmentation of the high energy nucleus}

There is a vast literature on different aspects of AA -
interactions. Important for our subject is the fact that complete
fragmentation of a projectile nucleus does not occur at
relativistic energies i.e. the nucleus is not split into
independent nucleons. The general feature of the independence of
the fragmentation process on the collision energy which is
observed at energies above 20-30 GeV is known as {\em limiting
fragmentation} \cite{Adam1,Back}. Some authors claim that the yield of heavy
fragments even rises with energy up to 200 GeV/nucleon due to an
increasing role of the electromagnetic excitation and dissociation
of projectile nuclei \cite{Wadd}. For instance, in the
interactions of Fe group nuclei (~22$<Z<$28~) with nuclear
emulsions at 20-65 GeV/nucleon such a heavy fragment as Sc
(~$Z$=21~) has been observed, amongst others \cite{Burn}.
Electromagnetic nuclear excitation and multifragmentation is
observed in ultra-peripheral collisions even at RHIC energies
(~$\sqrt{s}$ = 200 GeV/nucleon, which corresponds to an equivalent
energy of 4.3 PeV for gold in the laboratory coordinate system
\cite{Klei}~) and used as a collider luminosity monitor
\cite{Adle}. In what follows we postulate that the process of
nuclear fragmentation persists up to PeV energies in
AA-interactions.

\subsection{Rotation, polarization and deformation of the fragment}

Nuclear fragments arising in grazing peripheral collisions at
sub-GeV energies often rotate like a {\em 'sling'} and achieve a
high spin. This spin is polarized in the plane transverse to the
momentum of the projectile nucleus \cite{Fick}. The effect is used at low and
intermediate energies for the formation of polarized beams of low
intensity \cite{Satc}. In spite of the fact that the nuclear
fragment is a composite, and often excited, object it demonstrates
the features of collective motion and behaves like a liquid drop.
Some features of collective motion of nuclear matter, eg the
directed and elliptic flows and high transverse momenta of the
fragments, are preserved even at relativistic energies
\cite{Klei,Adam2}. In this connection we assume that the features of the
fragment's rotation and polarization persist at least up to PeV energies. 
An indirect indication that this assumption can be true is the alignment of 
energetic sub-cascades in multicore gamma-families detected by X-ray emulsion 
chambers at mountain altitudes and in the stratosphere \cite{Capd}. Such 
alignment can be expected when the excited and spinning nuclear fragment decays 
and emits high energy particles preferentially in its rotation plane, like splashes 
from a rotating wheel or sparks from a grindstone \cite{Muha}.   

Although the liquid drop model is not able to give a quantitative explanation of such 
fine features as gamma-ray spectra of excited nuclei, it is still useful for the 
modeling of the gross features, such as the shape of the excited nucleus, fluctuations 
of the collisional cross-section, the number of wounded nucleons etc (~see \S 3~).
The use of more sophisticated quantum-mechanical models such as the shell model can 
reveal some tiny effects, but for our purposes of a semi-quantative classical 
consideration the use of the liquid drop model is sufficient. 
  
Due to the rotation of the object with liquid drop properties its
shape is deformed. In the classical consideration this deformation is caused by 
centrifugal forces, in the quantum mechanics the deformation is attributed to the state
with a high angular momentum. The spherical fragment becomes an oblate spheroid with 
a shorter axis coincident with the axis of the spin. Following 
\cite{Drem} we call the effect of the rotation,
polarization and deformation of a nuclear fragment the {\em 'sling
effect'}. In the experimental study of deformed nuclei, spins of
up to a few tens  and deformations of nuclear sizes up to
$\sim$30\% have been observed \cite{Royer,Lafos}.

\section{A geometrical approach}
\subsection{General Remarks}

Many features of AA-interactions are considered within a purely
geometrical approach. The popular Bradt-Peters formula for the
interaction cross-section of two nuclei \cite{Brad} is just the
result of the geometrical examination of the collision between two
spheroids. The formulae for the mean number of
wounded nucleons in colliding nuclei were also derived using
geometrical arguments \cite{Bial,Shab}. We continue our consideration 
using the same geometrical approach.
As an example, we consider the collision of an iron nucleus
$^{56}$Fe, which is one of major constituents of cosmic rays at 
the energies in question here, with
a nitrogen nucleus $^{14}$N of the Earth's atmosphere. As a result
of this collision the fragment $^{45}$Sc emerges and the remaining
11 nucleons are liberated from the parent iron nucleus with, on
average, 8 being 'wounded'.

Some remarks are necessary also about cross-sections for nucleus-nucleus
collisions. In the (~assumed~) absence of coherent effects and 'shadowing'
phenomena, the number of nucleon-nucleon collisions will be independent 
of the manner in which the nucleons are packed, or distributed; thus, although
the mean cross-section for nucleus-nucleus interaction will depend on the shape 
of the nucleus the product of cross-section and number of 'wounded' 
nucleons will be a constant.  However,
the magnitude of the fluctuations about the mean \underline{will} depend 
on the manner of distribution. Clearly for the situation where the fluctuations
are large there is the possibility of misidentification of the primary mass
depending on whether the cross-section is high (~high mass~) or low 
(~great penetration and therefore light mass~). The latter situation is of
great importance here.

\subsection{The reduction of the mean cross-section}

Returning to the interaction of an iron nucleus with a nitrogen nucleus and an emerging
Sc fragment we assume that it moves down in the vertical
direction and is subject to the 'sling effect'. This means that it
rotates and has an ellipsoidal shape with a shorter axis
orientated in the horizontal direction. Let us denote its longer
axis as {\em 'a'} and the shorter axis as {\em 'c'}. The
deformation of the spheroid is often characterized by an
ellipticity $\epsilon$, which is defined as $\epsilon =
\sqrt{1-c^2/a^2}$. Due to the non-compressibility of nuclear
matter the volume of the deformed fragment should be equal to the
volume of the spherical one, i.e. $\frac{4}{3}\pi a^2 c =
\frac{4}{3}\pi R^3$, where $R$ is the radius of the spherical
fragment. Combining the formulae for $\epsilon$ and for the volume
we obtain $a=R/(1-\epsilon^2)^{1/6}$, $c=R(1-\epsilon^2)^{1/3}$.
If the ellipsoidal fragment moves down and its shorter axis is
orientated horizontally then its geometrical cross-section seen
'edge-on' from below is $\pi ac$. It means that this cross-section
is equal to $\pi R^2 (1-\epsilon^2)^{1/6}$, i.e. reduced compared
with a spherical fragment by the same factor of
$(1-\epsilon^2)^{1/6}$ as the increase of the longer axis.

 The prolate configuration of the spheroid in which the spin is oriented along the 
longer axis {\em 'a'} if it exists in nuclear collisions should have an opposite 
effect, increasing the mean cross-section, but we consider it unlikely due to the 
effect of centrifugal forces. In both cases the surface of the deformed fragment has to
 increase with the rising ellipticity by 0.4\% for $\epsilon$ = 0.5 and by 14\% for
$\epsilon$ = 0.9, but the volume according to our assumption remains constant. 

In what follows we examine two cases with $\epsilon = 0.5$ and
$0.9$. The latter corresponds to the increase of the larger
axis by about 32\% and to the reduction of the shorter axis by
43\%, which approximately corresponds to the magnitude of the
deformation observed in experiments hitherto. For these values of
$\epsilon$ the reduction of the purely geometrical cross-section
is 5\% and 24\% respectively, which definitely  should reduce the
mean interaction cross-section between this fragment and the
target nitrogen nucleus. The maximum observed hyperdeformation \cite{Lafos}
in which the ratio of a long to a short radius achieved 3:1 corresponds 
to an ellipticity $\epsilon \approx 0.94$. At the end of the paper we examine 
the hypothetical case of $\epsilon = 0.99$ to show the rise of the 
sling effect if the nuclear deformation grows with energy.  

In order to estimate the reduction of the cross-section we introduce the so called 
{\em collisional cross-section} $\sigma$ as a parameter for a single collision. It is 
equal to the overlap area between two colliding objects (~Figure 1~).
\begin{figure}[htb]
\begin{center}
\includegraphics[width=10cm,height=9.5cm,angle=-90]{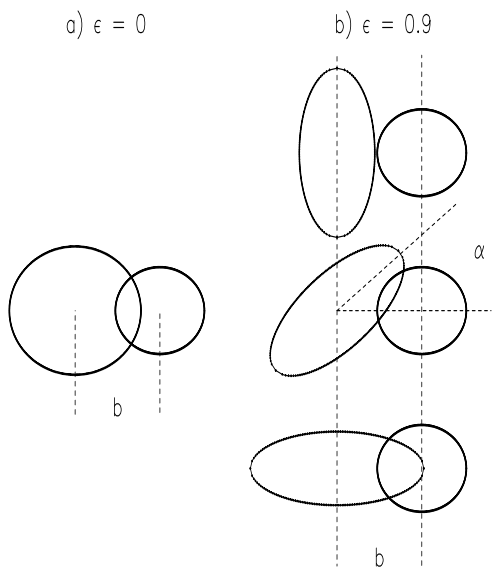}
\caption{\footnotesize A schematic view of the collision: a)
between two spherical nuclei and b) between an ellipsoidal
fragment and spherical target, the impact parameter being the same
in the two cases. It is seen that the overlap area in the latter
case depends on the orientation angle $\alpha$ between the larger
axis of the ellipsoid and the vector connecting the centers of the
two nuclei even for a fixed impact parameter {\em b}.}
\end{center}
\label{fig:sling1}
\end{figure}
In the case of the two spherical objects it depends just on the
radii of the objects $R,r$ and the impact parameter $b$. In the
case of the colliding ellipsoid and spheroid the overlap area
depends not only on $a,c,r$ and $b$, but also on the orientation
angle $\alpha$ between the longer axis of the ellipsoid and the
vector connecting the centers of the two objects \cite{Fick} (~see Figure 1~).
We have calculated this dependence for colliding nuclei of $^{45}$Sc
with $\epsilon = 0, 0.5$ and $0.9$ and $^{14}$N. The result of
the integration of the overlap area over the impact parameter $b$
as a function of the orientation angle $\alpha$ is shown in Figure
2. For simplicity the radii of non-deformed Sc and N nuclei have
been taken as $R,fm = 1.2\cdot A^{1/3}$, where A is the mass of
the nucleus and the interaction occured in every case when the
overlap area was non-zero. It is seen that the dependence of the
cross-section on $\alpha$ and the reduction of its mean value
compared with the case of the non-deformed fragment with $\epsilon
=0$ becomes stronger with an increase of the ellipticity.
\begin{figure}[htb]
\begin{center}
\includegraphics[width=10cm,height=15cm,angle=-90]{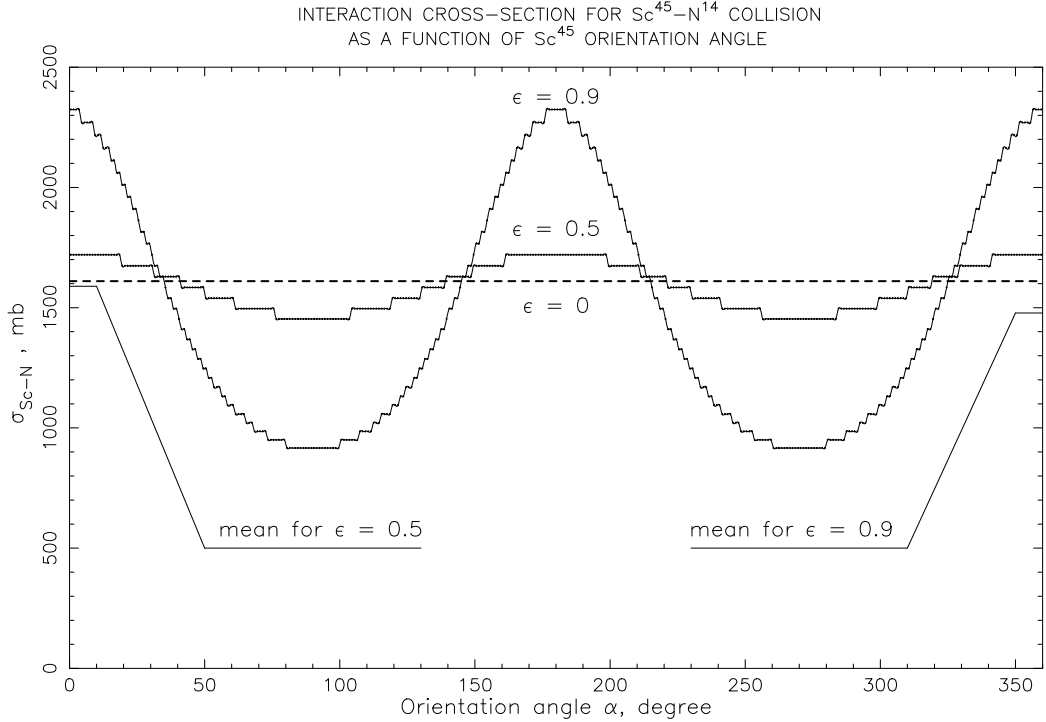}
\caption{\footnotesize The interaction cross-section for $^{45}$Sc
- $^{14}$N collision as a function of  $^{45}$Sc orientation angle
$\alpha$ for the ellipticity $\epsilon = 0$ (dashed line), $0.5$
and $0.9$ (~indicated in the Figure~). The mean values of the
cross-section averaged over the uniform distribution of $\alpha$
are also indicated for $\epsilon = 0.5$ and $0.9$. The reduction
of the mean cross-section for a deformed fragment of Sc with
respect to a spherical one with $\epsilon = 0$ is clearly seen.}
\end{center}
\label{fig:sig}
\end{figure}

\subsection{The attenuation of the deformed fragment}

When the collisional cross section fluctuates, the interaction rate of the fragment
changes and becomes non-exponential. We calculate it as
\begin{equation}
\frac{dP}{dz} = \int_0^{\infty} \sigma exp(-\sigma z) \frac{dP}{d\sigma} d\sigma
\end{equation}
where $P$ is the probability for either the fragment to interact
at the depth $z$, in the expression for $\frac{dP}{dz}$, or to
collide with a cross-section $\sigma$ in the expression for
$\frac{dP}{d\sigma}$. We used a gaussian approximation of the
$\sigma(\alpha)$ function in the range of $\alpha = 0^\circ -
90^\circ$ and the result of the integration compared with the
interaction rate for the spherical fragment is shown in Figure 3.
The non-exponential character of the interaction rate increases
with ellipticity.
\begin{figure}[htb]
\begin{center}
\includegraphics[width=10cm,height=15cm,angle=-90]{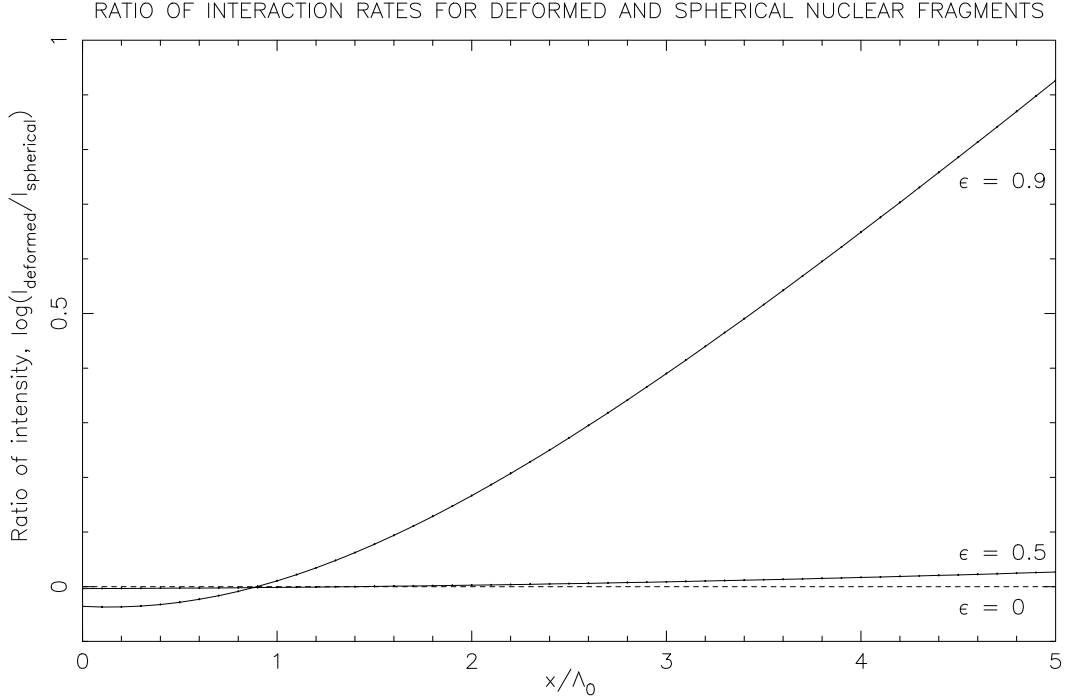}
\caption{\footnotesize The ratio of the interaction rate for
deformed (~$\epsilon$ = 0.5 and 0.9~) and spherical (~$\epsilon$ =
0~) Sc fragments in the air. $\Lambda_0$ is the mean
interaction length of the spherical Sc fragment in the air. The
reduced interaction rate at small depths leads to a reduced
attenuation and to a higher probability of finding the fragment
deep in the atmosphere.}
\end{center}
\label{fig:att_exp}
\end{figure}
It is seen that the ratio of interaction rates changes with
atmospheric depth. The reduced rate at small depths leads to a
higher probability for the fragment to penetrate deep into the
atmosphere. It is the effect which we need in order
 to increase the penetrating ability of atmospheric cascades.

\subsection{The number of wounded nucleons}

The attenuation of the atmospheric cascade induced by a nucleus is
determined not only by the interaction rate of the fragment, but
also by the number of nucleons participating in the interaction,
which are usually called 'wounded nucleons'. The mean number of wounded
nucleons and their fluctuations can be calculated using the Glauber approach,
but, since it gives the same result [23] as the geometrical approach, we use the 
latter as more visual and simple. The meaning of the
geometrical formulae used for the estimate of the mean number of
wounded nucleons in the projectile nucleus {\bf A} and in the
target nucleus {\bf B} can be seen from:
\begin{eqnarray}
n_W^A = A\frac{\bar{\sigma}_{pB}}{\bar{\sigma}_{AB}} = A\frac{S_B S_A}{\bar{\sigma}_{AB} S_A} = A\frac{S_{overlap}}{S_A}  \nonumber \\
n_W^B = B\frac{\bar{\sigma}_{pA}}{\bar{\sigma}_{AB}} = B\frac{S_A S_B}{\bar{\sigma}_{AB} S_B} = B\frac{S_{overlap}}{S_B}
\end{eqnarray}
where $n_W^A$ and $n_W^B$ are the mean numbers of nucleons in the {\bf A} or {\bf B} 
nucleus uniformly distributed over its geometrical cross-section, which is cut by an 
overlap area of colliding nuclei. In the formula (2) $\bar{\sigma}_{pA}$, 
$\bar{\sigma}_{pB}$ and $\bar{\sigma}_{AB}$
are ordinary mean cross-sections for the interaction of protons with {\bf A} and 
{\bf B} nuclei respectively and between {\bf A} and {\bf B} nuclei themselves, $S_A$,
$S_B$ are areas of {\bf A} and {\bf B} in the geometrical approach and $S_{overlap}$ is
described below. In \cite{EW3} we used this geometrical two-dimensional approach for 
the determination of the number of nucleons wounded in a {\em single} collision. 
In this case $S_{overlap}$ is the overlapping area which appears in one collision and
can be different in another one. It depends on the ellipticity of the fragment
$\epsilon$, impact parameter $b$ and orientation angle $\alpha$.

In this paper we use a more accurate three-dimensional approach, taking
\begin{eqnarray}
n_W^A = A\frac{V_A^{overlap}}{V_A} \nonumber \\
n_W^B = B\frac{V_B^{overlap}}{V_B}
\end{eqnarray}
where $V_A^{overlap}, V_B^{overlap}$ are the volume of the nucleus {\bf A} and {\bf B}
respectively cut by the overlapping part of the counterpart nucleus {\em (~tube~)} and 
$V_A, V_B$ are the total volumes of these nuclei. This approach, which could be called
{\em 'tube'} approach, implies that the nuclear density distribution inside the nucleus
and therefore along the tube is uniform.

The dependence of $n_w$ on $\alpha$ for $\epsilon$ = 0, 0.5 and 0.9 for different fixed
 impact parameters $b$ is shown in Figure 4a.
\begin{figure}[htb]
\begin{center}
\includegraphics[width=7.5cm]{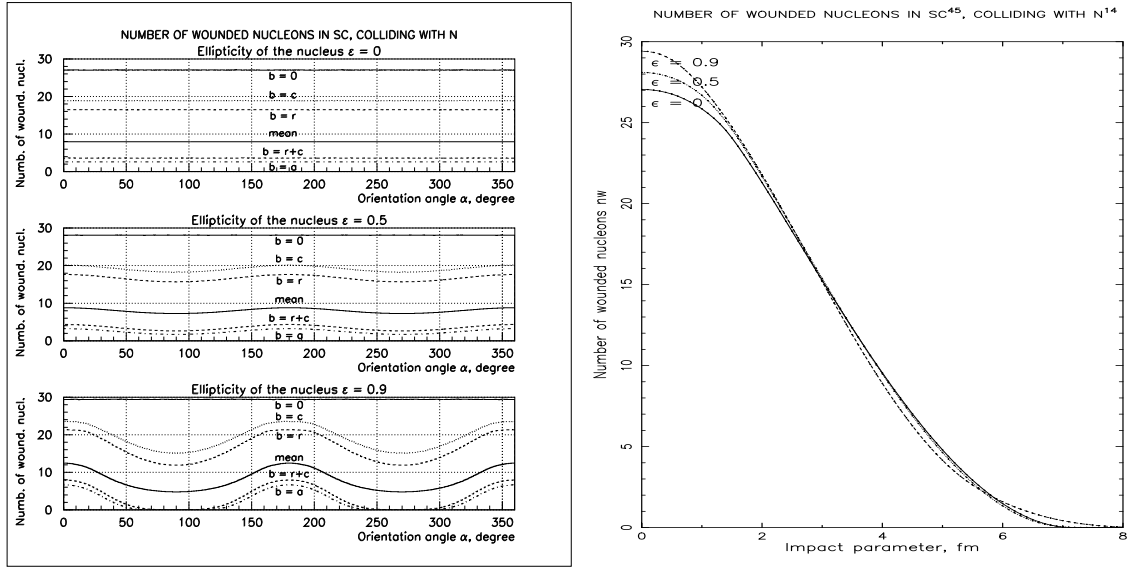}
\includegraphics[height=7.5cm,width=8cm]{sling4b.eps}
\caption{\footnotesize a) The dependence of the number of wounded nucleons $n_w$ in a
single collision on the orientation angle $\alpha$ for the ellipticity of the
projectile fragment $\epsilon$ = 0, 0.5 and 0.9 and for the fixed impact parameters
$b = 0, r$ - the radius of target nucleus, $c$ and $a$ - small and large axes of the
ellipsoid respectively, the sum $r+c$ and the
mean $\langle n_w (\alpha) \rangle$ averaged over the impact parameter $b$.
b) The dependence of the mean number of wounded nucleons
$\langle n_w (b) \rangle $ on the impact parameter $b$
averaged over the uniform distribution of orientation angles $\alpha$.}
\end{center}
\label{fig:nw}
\end{figure}

It is seen that the dependence on $\alpha$ increases with the ellipticity $\epsilon$.
With an increased ellipticity the maximum impact parameter at which the interaction
could occur increases too as well as the maximum length of the tube in the projectile 
nucleus. Due to this deformation and the constant nuclear density the range of the 
impact parameters and the maximum number of wounded nucleons increase too 
(~Figure 4b~).

\subsection{The distribution of the number of wounded nucleons and its correlation
with the cross-section}

Due to the stochastic fluctuations of the impact parameters and orientation angles
the number of wounded nucleons also fluctuates. Its distribution is shown in Figure 5a.
\begin{figure}[htb]
\begin{center}
\includegraphics[width=7.5cm]{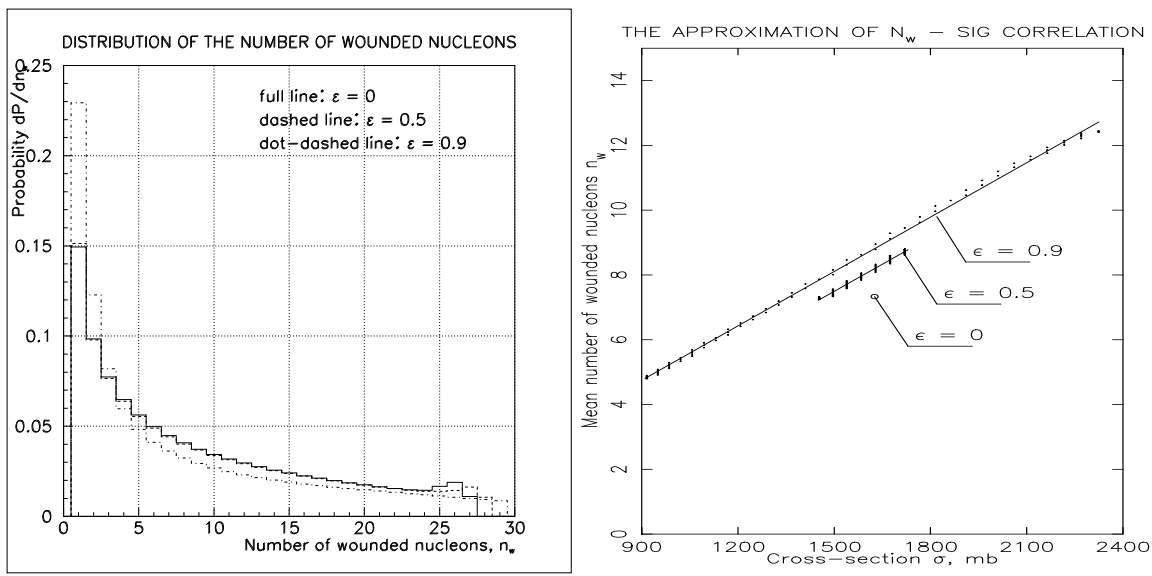}
\includegraphics[height=7.5cm,width=8cm]{sling5b.eps}
\caption{\footnotesize (a)-the distribution of the number of wounded nucleons  and
(b)-the correlation between the collisional cross-section and the number of wounded
nucleons for the different ellipticity of the projectile fragment: $\epsilon$ = 0, 0.5
and 0.9. Lines are linear fits of the numerical data.}
\end{center}
\label{fig:fluc}
\end{figure}
It is seen that the width of the distribution increases with the
ellipticity. An important consequence of the deformation is the
strong correlation between the number of wounded nucleons and the collisional 
cross-section. It is an evident feature
because $n_w$ is strongly connected with the overlap area $\sigma$ in the single
collision. This correlation is shown in Figure 5b for $n_w$ and $\sigma$ averaged over
all impact parameters. The existence of this correlation is a new feature for the
interaction of deformed fragments and is important for their propagation through an
absorber and the subsequent development of cascades initiated by high energy
 nuclei.

\section{The longitudinal development of the nucleus-induced cascade}

\subsection{The longitudinal development of the nucleon component}

In our example of the primary $^{56}$Fe induced cascade we adopted
the case where the interaction
 of the primary iron was normal with a deformed spinning $^{45}$Sc fragment emitted.
In the collision of the $^{45}$Sc fragment with a nitrogen nucleus
of the air on average 7 projectile nucleons were wounded and the
other 38 nucleons of the fragment liberated as spectators and
propagated through the atmosphere in an ordinary way. Therefore
the difference between the longitudinal development of the nucleon
component in an ordinary cascade and the cascade with a spinning
fragment is determined by the greater penetrating ability of the
fragment. We have calculated the interaction rate of nucleons in
our Fe-induced cascade with a spinning Sc fragment and compared it
with the nucleon interaction rate in the ordinary Fe-induced
cascade. The result for the case of $\epsilon$ = 0.9 is shown in
Figure 6.
\begin{figure}[htb]
\begin{center}
\includegraphics[height=10cm,width=10cm]{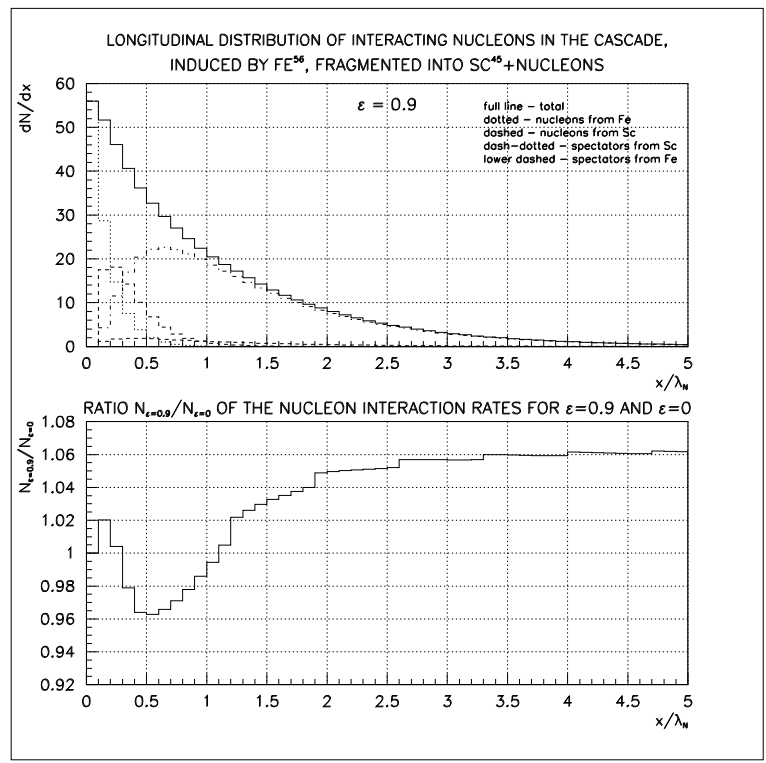}
\caption{\footnotesize a) The interaction rate of nucleons in Fe-induced cascade: 
dotted line - wounded nucleons from Fe, dashed line - wounded nucleons from Sc, 
dash-dotted line - spectators from Sc, lower dashed line - spectators from Fe.
b) The ratio of nucleon interaction rates in the cascade with a spinning Sc fragment
and in an ordinary Fe-induced cascade with non-spinning Sc.}
\end{center}
\label{fig:aa_fl}
\end{figure}

It is seen that in the cascade with a spinning fragment the
interaction rate decreases at the initial stages of the cascade
development and on the contrary increases later - the effect which
we needed to make the cascade able to penetrate deeper into
the atmosphere, although the absolute value of the effect does not
exceed a few percent even for $\epsilon$ = 0.9. For $\epsilon$ =
0.5 it is even less.

\subsection{The longitudinal development of the Fe-induced cascade}

In order to calculate the influence of the sling effect on the
development of the electromagnetic component of EAS we used the
approximate formulae for the cascades initiated by nucleons proposed in \cite{Cata} 
and used by us in \cite{EW2}. These formulae are results of the analytical solution of 
the system of kinetic equations for the development of the 
nucleon, pion and electromagnetic components of EAS \cite{EW2}. Interaction
rates of nucleons in the Fe-induced cascade were taken from the
previous subsection. The result for a primary Fe nucleus of 1 PeV
energy is shown in Figure 7.
\begin{figure}[htb]
\begin{center}
\includegraphics[height=10cm,width=10cm,angle=-90]{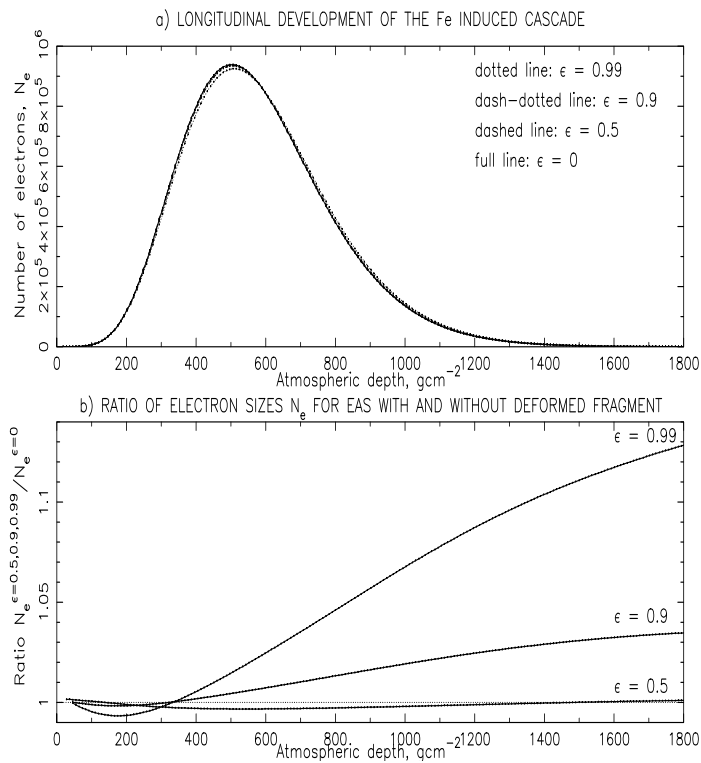}
\caption{\footnotesize a) The longitudinal development of 1 PeV Fe-induced EAS with
the emission of a Sc fragment with different ellipticity. The difference between
$\epsilon$ = 0, 0.5 and 0.9 is small and not seen at this graph. The difference between
$\epsilon$ = 0.99 and 0 is more pronounced: the cascade with a deformed fragment is 
shifted to a larger depth. b) The ratio of the
shower size $N_e$ for $\epsilon$ = 0.5, 0.9, 0.99 and 0 as the function of the 
atmospheric depth.}
\end{center}
\label{fig:emcas}
\end{figure}
It is seen that the development of the electromagnetic component resembless that of
the nucleon component, viz. the cascade develops later, its size at the early stages of
the development is smaller, but larger at large atmospheric depths. For moderate
ellipticities, not exceeding 0.9, the magnitude of the sling effect is small, however.
 The shift of $X_{max}$ - the maximum cascade development, does not exceed 
1 gcm$^{-2}$. The increase of the size at sea level for $\epsilon$ = 0.9 is about 
2\%. 

\subsection{The possible role of the sling effect at ultra-high energies}
 
The sling effect becomes stronger when the smaller radius of the deformed
fragment becomes comparable or even less than the radius of the target nucleus.
In our example of colliding $^{45}$Sc and $^{14}$N it happens just when the ellipticity
approaches 0.9. The maximum deformation observed hitherto corresponds to an 
ellipticity $\epsilon = 0.94$ \cite{Lafos}. However, these observations have been made
at low energies. We have assumed that at higher energies the deformation can be even 
higher 
and calculated the longitudinal development of the Fe 
induced cascade for the hypothetical deformation of a Sc fragment equal to 0.99. It is 
also shown in Figure 7. The sling effect for this case is certainly greater. The shift 
of $X_{max}$ increases up to 5 gcm$^{-2}$ and the increase of the shower size at the 
sea level grows by up to 7\%. Therefore the sling effect can be important if the 
deformation of fragments emitted in nucleus-nucleus collisions grows with
energy beyond the values observed hitherto.

The elongation rate for nucleus-induced cascades will be, in this case, also higher 
than without the sling effect and the interpretation of experimental
data on $X_{max}$ distributions should be re-examined; the effect leads to a higher
abundance of heavy nuclei in PeV cosmic rays.  
At the ultra-high energies ($\simeq 10^{20}$ eV) the effect can be very large and the 
consequences for the mass composition of the important extragalactic particles rather
profound.

In particular, the substantial increase of fluctuations in the collisional 
cross-section and the number of wounded nucleons could be responsible, at least partly,
for the seeming contradiction between the mean and the width of the $X_{max}$ 
distribution in the stereo Fly's Eye measurements \cite{Cass}. The mean 
$\langle X_{max} \rangle$ was even less than that expected for pure iron, but the 
fluctuations of $X_{max}$ exceeded those expected for pure protons and corresponded 
rather to the mixed composition.  

The relativistic expansion of the time needed for the formation of a stable rotation 
and deformation of the fragment cannot change the effect, since this time is about
10$^{-22}$ sec in the fragment's rest system and even such Lorentz-factors as 
$\sim$10$^{11}$ cannot make it comparable with the time needed to move between the 
collision in which this fragment has been produced, and its next collision, which is 
about 10$^{-5}$ sec at high atmospheric altitudes.

\section{Conclusions}

We have examined the possible slowing down of the development of
the atmospheric cascade initiated by a primary nucleus of high
energy due to the 'sling effect', which is the rotation, deformation
and polarization of nuclear fragments emitted in the
nucleus-nucleus interactions. At the moment it is difficult to
make accurate estimates of this effect, because of lack of
experimental data on the fragmentation of nuclei and on the excitation, polarization 
and interactions of the emitted nuclear fragments but we have made an attempt. In
the examined example of the emission of just one $^{45}$Sc fragment from the 
interaction of a 1 PeV primary $^{56}$Fe nucleus, if the ellipticity of the fragment 
does not exceed 0.9, the effect is rather small. The shift of the cascade maximum does
not exceed 1 gcm$^{-2}$ and the increase of the shower size at sea level is about 2\%. 
These changes alone are not sufficient to eliminate inconsistencies in the 
interpretation of experimental data on EAS discussed in the Introduction. 

However, numerical values of the mass, multiplicity and ellipticity of the fragment in 
the examined example were taken from the experiments at GeV and TeV energies. If there 
is an energy dependence of these characteristics and the ellipticity rises with energy,
 as seems likely, the sling effect can be much stronger. For the 
hypothetical case of $\epsilon = 0.99$ the shift of $X_{max}$ increases up to
5 gcm$^{-2}$ and the shower size at the sea level grows up to 7\%. 

If this rise is real the role of the sling effect at ultra-high energies can be very 
important and should be taken into account in the models of nucleus-nucleus 
interactions. It can have a profound effect on the inferred primary mass - 
a quantity of considerable astrophysical significance.

Finally, attention can be given to the phenomenon of 'alignment'; this phenomenon has 
various interpretations \cite{Muha,Drem} and it 
would be promising to search for the sling effect in the azimuthal alignment of the 
secondary particles on the event by event basis directly in nucleus-nucleus 
interactions at RHIC and LHC. 
\vspace{0.5cm}

{\large{\bf Acknowledgments}}\\

We thank M.Baldo and G.I.Orlova for useful discussions, R.A.Mukhamedshin for sending us
some unpublished materials and anonymous referees for critical remarks and suggestions.
One of us (ADE) thanks The Royal Society for financial support.


\begin{thebibliography}{99}
\bibitem{Heck} Heck D. et al., 1998, FZKA Report Forschungszentrum Karlsruhe 6019
\bibitem{EW1}  Erlykin A.D., Wolfendale A.W., 1998, Astroparticle Physics, {\bf 9}, 213
\bibitem{Ant1} Antoni T. et al. 1999, J.Phys.G: Nucl.Part.Phys., {\bf 25 }, 2161
\bibitem{Ulr1} Roth M. et al. 2001, Proc. 27th ICRC, Hamburg, {\bf 1}, 88
\bibitem{Hau1} Haungs A. et al. 2003, Progr. Nucl. Part. Phys., {\bf 66}, 1145
\bibitem{Hoe1} H\"{o}randel J.R. 2003, J.Phys.G: Nucl.Part.Phys., {\bf 29}, 2439
\bibitem{Yako} Yakovlev V.I. 2003, Nucl. Phys. B (~Proc.Suppl.), {\bf 122}, 417
\bibitem{EW2}  Erlykin A.D., Wolfendale A.W., 2002, Astroparticle Physics, {\bf 18}, 151
\bibitem{Adam1} Adamovich M.I. et al., 1989, Phys. Rev. Lett., {\bf 62}, 2801
\bibitem{Back} Back B.B., 2003, Phys. Rev. Lett., {\bf 91}, 052303
\bibitem{Wadd} Waddington C.J. et al., 1990, Proc. 21st ICRC, Adelaide, {\bf 8}, 87
\bibitem{Burn} Burnett T. et al. 1987, Phys. Rev. {\bf D35}, 824
\bibitem{Klei} Klein S.R., 2003, Nucl. Phys. B (~Proc.Suppl.), {\bf 122}, 76
\bibitem{Adam2} Adamovich M.I. et al., 2004, Physics of Atomic Nuclei, {\bf 67}, 290
\bibitem{Adle} Adler C. et al. nucl-ex/0206004
\bibitem{Fick} Fick D., 1981, Ann. Rev. Nucl. Part. Sci., {\bf 31}, 53
\bibitem{Satc} Satchler G.R., Introduction to Nuclear Reactions, Macmillan Press Ltd.,
New York, 1980
\bibitem{Capd} Capdevielle J.N. and Slavatinsky S.A., 1999, Nucl. Phys. B (~Proc. 
Suppl.), {\bf 75A}, 12
\bibitem{Muha} Mukhamedshin R.A., 2001, Nucl. Phys.B (~Proc.Suppl.), {\bf 97}, 122  
\bibitem{Drem} Dremin I.M., Man'ko V.I. 1998, Nuovo Cimento, {\bf 111A}, 439
\bibitem{Royer} Royer G. and Haddad F., 1993, Phys. Rev. C, {\bf 47}, 1302
\bibitem{Lafos} Lafosse D.R. et al., 1995, Phys. Rev. Lett., {\bf 74}, 5186
\bibitem{Brad} Bradt H.L., Peters B., 1948, {\bf 74}, 1828
\bibitem{Bial} Bialas A. et al., 1976, Nucl. Phys. {\bf B111}, 461
\bibitem{Shab} Shabelsky Yu.M., 1979, Acta Phys. Polonica, {\bf B10}, 1049
\bibitem{EW3}  Erlykin A.D., Wolfendale A.W., 2004, Nucl. Phys. B (~Proc.Suppl.),
{\bf 136}, 282
\bibitem{Cata} Catalano O. et al, 2001, Proc. 27th ICRC, Hamburg, {\bf 2}, 498
\bibitem{Cass} Cassiday G.L. et al, 1990, Astrophys. J., {\bf 356}, 669
\end{thebibliography}
\end{document}